\documentclass[twocolumn, pra,showpacs,superscriptaddress]{revtex4}
\usepackage{graphicx}
\usepackage{dcolumn}
\usepackage{bm}
\usepackage{amsmath}

\begin{document}

\title{Density-functional theory of two-component Bose gases in one-dimensional harmonic traps}
\author{Yajiang Hao}
\affiliation{Department of Physics, University of Science and
Technology Beijing, Beijing 100083, China}
\author{Shu Chen}
\email{schen@aphy.iphy.ac.cn} \affiliation{Beijing National
Laboratory for Condensed Matter Physics, Institute of Physics,
Chinese Academy of Sciences, Beijing 100190, China}
\date{\today}

\begin{abstract}
We investigate the ground-state properties of two-component Bose
gases confined in one-dimensional harmonic traps in the scheme of
density-functional theory. The density-functional calculations
employ a Bethe-ansatz-based local-density approximation for the
correlation energy, which accounts for the correlation effect
properly in the full physical regime. For the binary Bose mixture
with spin-independent interaction, the homogeneous reference system
is exactly solvable by the Bethe-ansatz method. Within the
local-density approximation, we determine the density distribution
of each component and study its evolution from Bose distributions to
Fermi-like distribution with the increase in interaction. For the
binary mixture of Tonks-Girardeau gases with a tunable inter-species
repulsion, with a generalized Bose-Fermi transformation we show that
the Bose mixture can be mapped into a two-component Fermi gas, which
corresponds to exact soluble Yang-Gaudin model for the homogeneous
system. Based on the ground-state energy function of the Yang-Gaudin
model, the ground-state density distributions are calculated for
various inter-species interactions. It is shown that with the
increase in inter-species interaction, the system exhibits
composite-fermionization crossover.

\end{abstract}
 \maketitle

\section{Introduction}
Experimental realization of the atomic mixtures and progress in
manipulating cold atom systems have opened exciting new
possibilities to study many-body physics of low-dimensional quantum
gases beyond the mean-field theory. In comparison with the
single-component systems, mixtures of quantum degenerate atoms may
form novel quantum many-body systems with richer phase structures.
Particularly, two-component Bose gases have been under intensive
studies both experimentally \cite
{Myatt,Modugno,Erhard,exp08,Experiment} and theoretically \cite
{Ho96,Ao,Pu,Cazalilla,Zhou}. Due to the competition among the
inter-species and intra-species interactions, many interesting
phenomena such as phase separation, new quantum states and phase
transitions arise in two-component cold atomic gases
\cite{Ho96,Ao,Pu,Cazalilla,Zhou}. With strong anisotropic magnetic
trap or two-dimensional optical lattice the radial degrees of
freedom of cold atoms are frozen and the quantum gas is described by
an effective one-dimensional (1D) model
\cite{Stoeferle,Paredes,Toshiya}. As a textbook example the 1D
Tonks-Girardeau (TG) gas \cite{Girardeau} was observed firstly
\cite{Paredes,Toshiya}. Since then, not only the single-component
bosonic gas but also 1D multi-component atomic mixtures have become
the current research focus. In principle, both the intra-component
and inter-component interactions can be tuned via the magnetic
Feshbach resonance, which allows us to study the bosonic mixture in
the whole interaction regime. Very recently, two-species Bose gases
with tunable interaction have been successfully produced \cite
{exp08,Experiment}.

In general, 1D quantum systems exhibit fascinating physics
significantly different from its three-dimensional counterpart
because of the enhanced quantum fluctuations in 1D
\cite{Olshanii,Lieb,Olshanii2,Petrov,Dunjko,Chen}. The mean-field
theory generally works not well for the 1D systems except in very
weakly interacting regime.  When interaction is strong enough,
non-perturbation methods, for instance, Bose-Fermi mapping (BFM),
Bethe ansatz and Bosonization method, have to be exerted to properly
characterize the features of the system \cite
{Girardeau07,DeuretzbacherPRL,
Guan,LiYQ,HaoEPJD,TBCHao,Zoellner08,Cazalilla,Fuchs,Guan07}. For a
single-component Bose gas, it has been shown that the density
profiles continuously evolve from Gaussian-like distribution of
bosons to shell-structured distribution of fermions
\cite{Hao06,Deuretzbacher,Zoellner,Cederbaum} with the increase in
repulsion strength. Similar behavior of composite fermionization has
been observed in the 1D Bose-Bose mixtures
\cite{TBCHao,Zoellner08,HaoEPJD}. For two-component Bose mixture
with spin-independent repulsions, the normalized density
distribution of each component displays the same behavior up to a
normalized constant which is proportional to respective atomic
number of each component \cite{TBCHao}. While the extended
Bose-Fermi mappings \cite{Girardeau07,Guan,DeuretzbacherPRL} are
restricted to the infinitely repulsive limit, some sophisticate
methods, such as multi-configuration self-consistent Hartree method
\cite{Zoellner08}, numerical diagonalization method \cite{HaoEPJD},
and Bethe ansatz method \cite{LiYQ,Fuchs,TBCHao,Guan07}, have been
applied to study the two-component Bose gases in the whole repulsive
regime. Nevertheless, the above methods either suffer the small-size
restriction \cite{Zoellner08,HaoEPJD} or only limit to the
integrable models \cite{LiYQ,Fuchs,TBCHao,Guan07} which generally do
not cover the realistic system confined in an external harmonic
trap.

In this work, we use the basic idea of density functional theory
(DFT) to investigate the ground-state properties of the
two-component Bose gases trapped in harmonic potential in the full
interacting regime. It is well known that the DFT can handle the
system with large sizes and will not encounter the exponential wall
that the above traditional multiparticle wave-function methods
faced. The effects of the external potential will be accounted by
using the local-density approximation (LDA) \cite
{Dunjko,Kolomeisky,Ohberg,Zubarevboson,Zubarev,Gao06,Gao07,Gao08,HHu,Astrakharchik}.
The combination of DFT and LDA had been applied to deal with the
confined 1D single-component Bose gas \cite
{Dunjko,Kolomeisky,Ohberg,Zubarevboson} and Fermi gas \cite
{Zubarev,Gao06,Gao07,Gao08,HHu,Astrakharchik} successfully. However,
the two-component Bose gas has rarely been studied. The present work
will study the ground-state properties of the confined Bose mixture
and focus on two cases where the ground state energy function for
the homogeneous system can be obtained exactly. In the first case,
we consider the two-component Bose gas with spin-independent
interaction. In the absence of external potential, this model is
exactly solvable by the Bethe-ansatz method and thus the
ground-state energy density function can be extracted from the
Bethe-ansatz solution \cite{LiYQ,Guan07,TBCHao,Fuchs}. In the second
case, we consider the two-component Bose gas with infinitely
repulsive intra-component interactions and a tunable inter-component
interaction. We will show that such a model can be mapped into the
Yang-Gaudin model with generalized Bose-Fermi transformation and
thus the ground-state energy function can be obtained from the
Bethe-ansatz solution of the Yang-Gaudin model \cite{GaudinYang}. In
the scheme of DFT, the ground-state density profiles are obtained by
numerically solving the coupled nonlinear Schr\"{o}dinger equations
(NLEs) for both the equal-mixing Bose mixture and polarized Bose
mixture.

The present paper is organized as follows. Section II investigates
two-component Bose gas with spin-independent interaction and
introduce the method. Section III is devoted to the mixture of two
TG gases with tunable inter-species repulsion. A summary is given in
the last section.

\section{Binary Bose mixture with spin-independent interaction}

We consider a two-component Bose gas confined in the external
potential $ V_{ext}\left( x\right) $ independent of the species of
atoms, where the atomic mass of the $i$-component is $m_i$. The atom
numbers in each component are $N_1$ and $N_2$ and $N=N_1+N_2$ is the
total atom number. The many-body Hamiltonian in second quantized
form can be formulated as
\begin{eqnarray}
\mathcal{H} &=&\int dx\sum_{i=1,2}\left\{ \hat{\Psi}_i^{\dagger }(x) \left[ -%
\frac{\hbar ^2}{2m_i} \frac{\partial^2} {\partial x^2} +V_{ext}( x ) \right]
\hat{\Psi}_i(x) \right.  \nonumber \\
&&\left. +\frac{g_i}2\hat{\Psi}_i^{\dagger }(x)\hat{\Psi}_i^{\dagger }(x)%
\hat{\Psi}_i(x)\hat{\Psi}_i(x)\right\}  \nonumber \\
&&+g_{12}\int dx \hat{\Psi}_1^{\dagger }(x)\hat{\Psi}_2^{\dagger }(x)\hat{%
\Psi}_2(x)\hat{\Psi}_1(x),  \label{HamB}
\end{eqnarray}
in which $g_i$ ($i=1,2$) and $g_{12}$ denote the effective intra-
and inter-species interaction that can be controlled experimentally
by tuning the corresponding scattering lengthes $a_1$, $a_2$ and
$a_{12}$, respectively \cite{exp08,Experiment}. $\Psi _i^{\dagger
}(x)$ ($\Psi _i(x)$) is the field operator creating (annihilating)
an $i$-component atom at position $x$. Here we will consider the
two-component Bose gas composed of two internal states of same
species of atoms such that we have the same mass for all atoms
$m_1=m_2=m$. When the intra- and inter-component interactions are
equal ($g_1=g_2=g_{12}=g$), the system is governed by the
spin-independent Hamiltonian with the first quantized form:
\begin{equation}
H=\sum_{j=1}^N \left[ -\frac{\partial ^2}{\partial x_j^2} + V_{ext}(x_j)
\right] + 2c\sum_{j<l}\delta (x_j-x_l),
\end{equation}
where $c=mg/\hbar^2$. The model in the absence of external trap is
integrable and can be diagonalized via Bethe ansatz method
\cite{LiYQ,Guan07}.

We first focus on the homogeneous case by setting $V_{ext}(x) = 0$.
The effects of the external potential will be discussed latter in
the scheme of the LDA and DFT. Before the application of DFT based
on the Bethe-ansatz solution, we first give a brief summary of the
Bethe-ansatz solution for the integrable two-component Bose gas. For
the eigenstate with total spin $S=N/2-M$ ($0\leq M \leq N/2$), the
Bethe ansatz equations for the integrable two-component Bose gas
take the following form \cite{LiYQ,Guan07}
\begin{eqnarray}
& & k_jL =2 \pi I_j -\sum_{l=1}^N \tan ^{-1}\frac{k_j-k_l}c + \sum_{\alpha
=1}^{M} \tan ^{-1}\frac{k_j-\Lambda _\alpha }{c/2},  \nonumber \\
& & \sum_{j=1}^N \tan ^{-1}\frac{\Lambda _\alpha -k_j}{c/2} =2 \pi J_\alpha
+\sum_{\beta \neq \alpha }^{M} \tan ^{-1}\frac{\Lambda _\alpha -\Lambda
_\beta }c,  \label{BAE}
\end{eqnarray}
where $k_j$ and $\Lambda _\alpha $ are the quasi-momentum and spin rapidity,
respectively, the quantum numbers $I_j$ and $J_\alpha $ take integer or
half-integer values, depending on whether $N-M$ is odd or even. In term of
the quasi-momentums, the energy eigenspectrum is given by $E=\sum_{j=1}^N
k_j^2$. In the thermodynamic limit, i.e., $N,L\to \infty$ with $N/L$ and $%
M/L $ finite, if one introduces density of $k$ roots $\rho(k)$ and
density of $\Lambda$ roots $\sigma(k)$ as well as the corresponding
hole root densities $\rho^h(k)$ and $\sigma^h(\lambda)$, the
Bethe-ansatz equations thus is simplified to two coupled integral
equations
\begin{eqnarray}
\rho(k)+\rho^h(k)&=&\frac{1}{2\pi}+\frac{1}{2\pi}\int_{-Q}^{Q}\frac{2c
\rho(k^{\prime})}{c^2+(k-k^{\prime})^2}dk^{\prime}  \nonumber \\
& &-\frac{1}{2\pi} \int_{-B}^{B}\frac{c\sigma(\lambda)}{c^2/4+(k-\lambda)^2}%
d\lambda  \label{BAE2a} \\
\sigma(\lambda)+\sigma^h(\lambda)&=&\frac{1}{2\pi}\int_{-Q}^{Q}\frac{c
\rho(k)}{c^2/4+(\lambda-k)^2}dk  \nonumber \\
& & -\frac{1}{2\pi} \int_{-B}^{B}\frac{2c\sigma(\lambda)}{%
c^2+(\lambda-\lambda^{\prime})^2}d\lambda^{\prime}.  \label{BAE2b}
\end{eqnarray}
The integration limits $Q$ and $B$ are determined by $N/L=\rho=\int_{-Q}^Q%
\rho(k)dk$ and $M/L=\int_{-B}^{B}\sigma(\lambda)d\lambda$.

At zero temperature, the ground state corresponds to the case with
$M=0$ \cite{LiYQ,Guan07,Fuchs,TBCHao}, where $I_j=\left( N+1\right)
/2-j$ and $J_\alpha $ is an empty set. In other words, the ground
state corresponds to the configuration
$\sigma(\lambda)=\rho^h(k)=0$, i.e., no holes in the charge degrees
of freedom and no quasiparticles in the spin degrees of freedom.
Consequently, the ground states with $S=N/2$ are $\left( N+1\right)
$-fold degenerate isospin `ferromagnetic' states \cite
{LiYQ,Eisenberg,Yang,TBCHao}. Therefore the Bethe ansatz equation
for the ground state reduces to \cite{LiYQ,TBCHao}
\begin{eqnarray}
\rho(k) = \frac{1}{2\pi}+\frac{1}{2\pi}\int_{-Q}^{Q}\frac{2c \rho(k^{\prime})%
}{c^2+(k-k^{\prime})^2}dk^{\prime},  \nonumber
\end{eqnarray}
or 
\begin{eqnarray}
g(x) = \frac{1}{2\pi}+\frac{1}{2\pi}\int_{-1}^{1}\frac{2 \lambda
g(x^{\prime})}{\lambda ^2+(x-x^{\prime})^2}d x^{\prime},  \label{LL1}
\end{eqnarray}
where we have made the replacement of variables $k=Qx$, $c=Q\lambda$ and $%
g(x)=\rho(Q x)$ according to \cite{Lieb}. The ground-state energy
density can be expressed as
\begin{equation}
\epsilon \left( \rho \right) =\frac{\hbar ^2}{2m} \rho ^2 e\left( \gamma
\right)  \label{muhom}
\end{equation}
where $\rho =\sum_i\rho _i$ is the total density and
\begin{equation}
e\left( \gamma \right) = \frac{\gamma^3}{\lambda^3} \int_{-1}^{1} g(x) x^2 d
x  \label{LL2}
\end{equation}
with the dimensionless parameter $\gamma =\frac{c}{\rho }$. In terms of new
variables, the equation $\rho=\int_{-Q}^Q\rho(k)dk$ is rewritten as
\begin{equation}
\lambda =\gamma \int_{-1}^1 g(x) dx .  \label{LL3}
\end{equation}

The expression of ground state energy is identical to that of
Lieb-Liniger Bose gas \cite{Lieb} except that now $\rho$ should be
replaced by the total density. Here $e\left( \gamma \right) $ can be
obtained by numerically solving the integral equations (\ref{LL1}),
(\ref{LL2}) and (\ref{LL3}) \cite {Dunjko,Lieb} and we find that it
can be fitted very accurately by the rational function
\begin{equation}
e\left( \gamma \right) =\frac{\gamma \left( 1+p_1\gamma +p_2\gamma
^2 \pi ^2/3\right) }{1+q_1\gamma +q_2\gamma ^2+p_2\gamma ^3}
\label{fitfunc}
\end{equation}
with $p_1=2.05737$, $p_2=0.0688097$, $q_1=3.16861$ and $q_2=0.898297$. In
Fig. 1 we compare the fit function with the exact numerical result, where
the exact match between them is displayed. In the weakly interacting TF
regime ($\gamma \ll 1$) $e\left( \gamma \right) $ is approximated as $%
e\left( \gamma \right) =\gamma $ and in the strongly interacting TG regime ($%
\gamma \gg 1$) we have $e\left( \gamma \right) =\pi ^2/3$ such that in these
two limits the energy density take the form of
\[
\epsilon \left( \rho \right) =\{
\begin{array}{ll}
\frac{\hbar^2}{2m} c \rho = \frac{g}{2} \rho,  & \gamma \ll 1 \\
\rho ^2 \frac{\pi ^2\hbar ^2}{6m}, & \gamma \gg 1
\end{array}
.
\]
Alternatively, $e\left( \gamma \right) $ can also be formulated by
the
rational functional with only two parameters $p=-5.0489470$ and $%
q=-20.8604983$
\begin{equation}
e\left( \gamma \right) =\frac{\gamma \left( 1+p\gamma \pi ^2/3
\right) }{ 1+q\gamma +p\gamma ^2},  \label{fitfunc2p}
\end{equation}
which also match well with the exact numerical result according to Fig. 1
although the accuracy is less perfect than Eq. (\ref{fitfunc}).
\begin{figure}[tbp]
\includegraphics[width=3.5in]{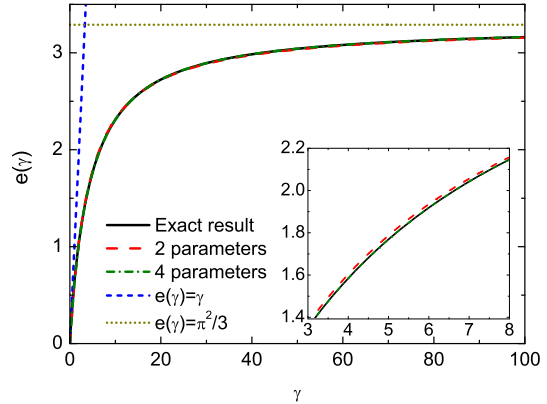}\newline
\caption{(color online) Fit function with four parameters and two
parameters. }
\label{fig1}
\end{figure}

In order to deduce the bulk properties of non-uniform systems, we
have to combine the analytical solution of the homogeneous system
with local density approximation. Within the local density
approximation, one can assume that the system is in local
equilibrium at each point $x$ in the external trap, with local
energy per particle provided by Eq. (\ref{muhom}). Therefore the
energy functional of an inhomogeneous system can be formulated as
\begin{equation}
\mathcal{E(\rho)}=\int dx\left[ \frac{\hbar ^2}{2m} \left| \frac{d}{dx}
\sqrt{\rho} \right|^2 + V_{ext}(x) \rho + \rho \epsilon \left( \rho \right)
\right]  \label{EF0}
\end{equation}
where $\epsilon(\rho )$ takes the form of Eq.(\ref{muhom}) which is
the exact ground-state energy density of the homogenous system \cite
{Dunjko,Minguzzi,Zubarevboson,Kolomeisky,Lieb2003}. The first
gradient term represents additional kinetic energy associated with
the inhomogeneity of the gas that is not accounted for by the
``local" kinetic energy included in the homogenous energy
\cite{Lieb2003,Ohberg}. Next we shall consider the case with
external potential being a weak harmonic trap $V_{ext}\left(
x\right) =\frac 12m\omega ^2x^2$. For the convenience of latter
calculation, we introduce $\Phi _i=\sqrt\rho_i$ and thus we have
$\rho= |\Phi_1|^2 + |\Phi_2|^2$. In terms of $\Phi_i$, the energy
functional can be represented as
\begin{equation}
\mathcal{E}=\int dx\left[ \sum_{i=1,2}\Phi _i^{*}\left( -\frac{\hbar ^2}{2m}%
\frac{d^2}{dx^2}+V_{ext}\right) \Phi _i+\rho \epsilon \left( \rho \right)
\right] .  \label{EF}
\end{equation}
The ground state corresponds to the minimization of the energy functional
Eq. (\ref{EF}). For the two-component bosonic system, the particle number of
each component is conserved. In order to obtain the ground state from a
global minimization of $\mathcal{E}$ with the constraints on both $N_i$, we
introduce Lagrange multiplier chemical potential $\mu _1$ to conserve $N_1$
and $\mu _2$ to conserve $N_2$. The ground state is then determined by a
minimization of the free-energy functional $\mathcal{F}=\mathcal{E}-\mu
_1N_1-\mu _2N_2$.
As shown in Ref. \cite{FDalfovo}, the minimization of energy functional
corresponds of the ground state of the modified coupled nonlinear
Schr\"{o}dinger equations
\begin{eqnarray}
\left( -\frac{\hbar ^2}{2m}\frac{d^2}{dx^2}+V_{ext}\right) \Phi _1+\tilde{F}%
\left( \rho \right) \Phi _1 &=&\mu _1\Phi _1,  \nonumber \\
\left( -\frac{\hbar ^2}{2m}\frac{d^2}{dx^2}+V_{ext}\right) \Phi _2+\tilde{F}%
\left( \rho \right) \Phi _2 &=&\mu _2\Phi _2.  \label{NLEs}
\end{eqnarray}
with the normalization condition $\int dx|\Phi _i(x)|^2=N_i$, where
\[
\tilde{F}\left( \rho \right)=\frac \partial {\partial \rho }\left[ \rho
\epsilon \left( \rho \right) \right] =\frac{\hbar ^2\rho ^2}{2m}\left[
3e\left( \gamma \right) -\gamma \partial _\gamma e\left( \gamma \right)
\right].
\]
In the weakly interacting regime ($\gamma \ll 1$) and strongly interacting
regime ($\gamma \gg 1$) the explicit forms are
\[
\tilde{F}\left( \rho \right) =\{
\begin{array}{ll}
\text{ \ \qquad }g \rho & \gamma \ll 1, \\
\pi ^2\hbar ^2\rho ^2/2m & \gamma \gg 1,
\end{array}
\]
while in the middle regime which can be obtained by Eq. (\ref{fitfunc}) or
Eq. (\ref{fitfunc2p}).

By numerically solving the coupled nonlinear Schr\"{o}dinger
equations Eq.(\ref{NLEs}) we can obtain the ground state density
profiles of each component in the full interacting regime, which are
displayed in Fig. 2 for the system with the atomic number $N_1=20$
and $N_2=15$. Here the dimensionless interacting constant
$U=g/l\hbar \omega$ and the length $x$ is in unit of $l=\sqrt{\hbar
/{m \omega}}$. In order to compare the distribution of each
component, both of them are normalized to one, {\it i.e.}, $n_i(x)=
\rho_i(x)/N_i$ where $\int \rho_i(x)dx=N_i$. It is shown that in the
full physical regime the density distribution of two components
match exactly. That is to say, the density distribution of
$i$-component fulfills a simple relation with the total distribution
\begin{equation}
\rho _i (x)=\frac{N_i }N\rho _{\text{tot}}(x)  \label{ruo}
\end{equation}
with $\rho _{\text{tot}}(x)=\rho _1 (x) + \rho _2 (x)$. Similar to
the case of single species of Bose gas, all of them exhibit the
evolution from Bose distribution to Fermi-like one with the increase
in interaction constant. In the weakly interacting regime the
density profiles display the Gauss-like Bose distribution, while in
the strongly interacting regime Bose atoms behave like fermions,
which distribute uniformly in a more extensive area and the density
profiles decrease to zero rapidly at the boundary. In order to
compare the results obtained from the density-functional
calculations to the exact results obtained by Bose-Fermi mapping
method, in Fig. 3 we show the density profiles of each component and
the total density profile in the strongly interacting limit for
$U=500$. Here each components are normalized to their particle
numbers and clearly they satisfy the relation of eq.(\ref{ruo}). The
total density profile obtained by BFM method is given by
$\rho_{TG}(x)= \sum_{n=0}^{N-1} |\phi_n(x)|^2  $ which is also shown
in Fig. 3. Here $\phi_n(x)$ denotes the eigenstate of the single
particle Hamiltonian of harmonic oscillator. It turns out that the
results obtained from both methods agree well with each other in the
full space except that the density profile obtained by BFM displays
oscillations. In the limit of large particle number, the differences
become imperceptible.
\begin{figure}[tbp]
\includegraphics[width=3.5in]{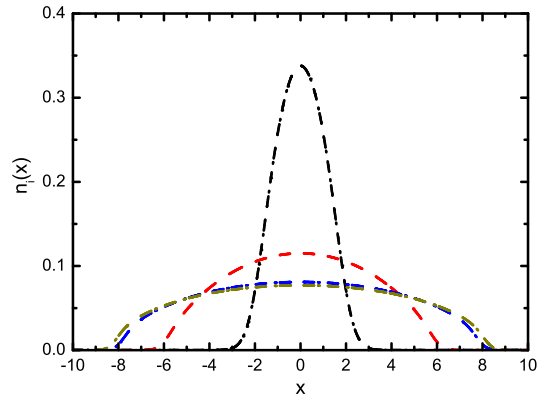}\newline
\caption{(color online) Normalized ground-state density distributions for $%
N_1=20$ (dashed lines) and $N_2=15$ (dotted lines). From the top to the
bottom: $U=0.2, 2.0,20,100$. Unit of $n_i(x)$: $N_i/l$; Unit of $x$: $l$.}
\label{fig2}
\end{figure}
\begin{figure}[tbp]
\includegraphics[width=3.2in]{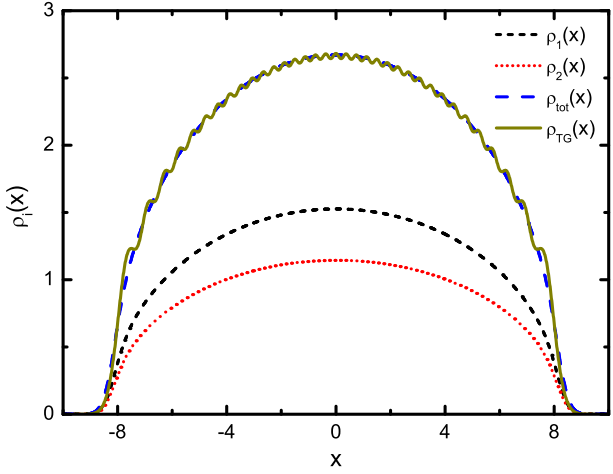}\newline
\caption{(color online) Ground-state density distributions for
$N_1=20$ (dashed lines) and $N_2=15$ (dotted lines) with $U=500$.
Unit of $\rho_i(x)$: $ 1/l$; Unit of $x$: $l$.} \label{fig3}
\end{figure}

\section{Mixture of TG gases with tunable inter-component interaction}
In this section, we focus on the mixture of two TG gases with
infinitely strong intra-species interaction, i.e., $g_i=\infty $
$(i=1,2)$ and a tunable inter-species interaction $g_{12}=g$. In
this limit, the original Hamiltonian Eq. (\ref{HamB}) is simplified
to
\begin{eqnarray}
\mathcal{H}^{TG}  &=&\int dx\sum_{i=1,2}\left\{
\hat{\Psi}_i^{\dagger }(x) \left[ - \frac{\hbar ^2}{2m_i}
\frac{\partial^2} {\partial x^2} +V_{ext}( x ) \right]
\hat{\Psi}_i(x) \right.  \nonumber \\
&&\left. + g  \hat{\Psi}_1^{\dagger }(x)\hat{\Psi}_2^{\dagger }(x)\hat{%
\Psi}_2(x)\hat{\Psi}_1(x) \right\} ,  \label{HamTG}
\end{eqnarray}
with the hard-core constraints $\hat{\Psi}_i^{\dagger }(x)\hat{\Psi}%
_i^{\dagger }(x)=\hat{\Psi}_i(x)\hat{\Psi}_i(x)=0$ and $\left\{ \hat{\Psi}%
_i(x),\hat{\Psi}_i^{\dagger }(x)\right\} =1$. For the mixture of TG
gases, it is convenient to introduce the generalized Bose-Fermi
transformations \cite{JWTSite}
\begin{eqnarray}
\hat{\Psi}_1(x) &=&\exp \left[ i\pi \int_{-\infty }^x n^F_{1} \left(
z\right) dz\right] \Psi^F_{1} \left( x\right)
\\
\hat{\Psi}_2(x) &=&\exp \left[ i\pi \int_{-\infty }^\infty n^F_{1}
\left( z\right) dz\right]  \nonumber \\
&&\times \exp \left[ i\pi \int_{-\infty }^x  n^F_2 \left( z\right)
 dz\right] \Psi _2^F\left( x\right) \label{JW2}
\end{eqnarray}
where $\Psi _i^{F\dagger }\left( x\right) $ and $\Psi _i^F\left(
x\right) $ are the creation and annihilation operator at location
$x$ for $i$-component fermions and $n_i^F(x)=\Psi _i^{F\dagger
}\left( x\right) \Psi _i^F\left( x\right) $. With the above
transformations, the Hamiltonian (\ref{HamTG}) is mapped into the
two-component fermionic gas with the Hamiltonian given by
\begin{eqnarray}
\mathcal{H}^{F}  &=&\int dx\sum_{i=1,2}\left\{
\hat{\Psi}_i^{F\dagger } \left[ - \frac{\hbar ^2}{2m_i}
\frac{\partial^2} {\partial x^2} +V_{ext}( x ) \right]
\hat{\Psi}_i^F \right.  \nonumber \\
&&\left. + g \hat{\Psi}_1^{F\dagger}
(x)\hat{\Psi}_2^{F\dagger}(x)\hat{\Psi}_2^F(x)\hat{\Psi}_1^F(x)
\right\} .  \label{HamF}
\end{eqnarray}
The first term on the right side of eq. (\ref{JW2}) is introduced to
enforce the fermionic annihilation operators $\hat{\Psi}_1^F(x)$ and
$\hat{\Psi}_2^F(x)$ fulfilling the anti-commutation relations $\{
\hat{\Psi}_1^F(x), \hat{\Psi}_2^F(x)\}=0$.

Since the bosonic model of (\ref{HamTG}) is related to the fermionic
model of (\ref{HamF}) by an unitary transformation, they have the
same energy spectrum. Furthermore the ground-state properties which
are independent of the statistical properties of bosonic system, for
example, the ground-state density distribution, can be obtained by
consulting for the corresponding fermionic model. Obviously the
homogeneous system of (\ref{HamF}) with $ V_{ext}\left( x \right)
=0$ is the Yang-Gaudin model which can be solved exactly by means of
the Bethe ansatz method \cite{GaudinYang}. The Bethe-ansatz
equations take the form of
\begin{eqnarray}
k_jL &=&2\pi I_j-2\sum_{\alpha =1}^M\tan ^{-1}\frac{k_j-\Lambda
_\alpha }{c/2
}  \nonumber \\
\sum_{j=1}^N\tan ^{-1}\frac{\Lambda _\alpha -k_j}{c/2} &=&\pi J_\alpha
+\sum_{\beta =1}^M\tan ^{-1}\frac{\Lambda _\alpha -\Lambda _\beta }c
\label{BAETG}
\end{eqnarray}
for the quasi-momentum $\left\{ k_j\right\} $ and rapidity $\left\{ \Lambda
_\alpha \right\} $ with the quantum number $I_j$ and $J_\alpha $. While in
the thermodynamic limit the equations (\ref{BAETG}) reduce to two coupled
integral equations
\begin{eqnarray}
\rho \left( k\right)  &=&\frac 1\pi \int_{-B}^Bd\Lambda
\frac{c^{\prime }\sigma \left( \Lambda \right) }{c^{\prime 2}+\left(
k-\Lambda \right) ^2}
+\frac 1{2\pi },  \nonumber \\
\sigma \left( \Lambda \right)  &=&-\frac 1\pi \int_{-B}^Bd\Lambda
^{\prime } \frac{c\sigma \left( \Lambda ^{\prime }\right)
}{c^2+\left( \Lambda -\Lambda
^{\prime }\right) ^2}  \nonumber \\
&&+\frac 1\pi \int_{-Q}^Qdk\frac{c^{\prime }\rho \left( k\right)
}{c^{\prime 2}+\left( \Lambda -k\right) ^2}  \label{BAETGIng}
\end{eqnarray}
with $c'=c/2$, where $Q$ and $B$ are determined by
$N/L=\int_{-Q}^Q\rho (k)dk$ and $M/L=\int_{-B}^B\sigma (\lambda
)d\lambda $. If we make the replacement of variables $x=k/Q$,
$y=\Lambda /B$, $g_c\left( x\right) =\rho \left( xQ\right) =\rho
\left( k\right) $ and $ g_s\left( y\right) =\sigma \left( yB\right)
=\sigma \left( \Lambda \right) $ , the above integral equations
(\ref{BAETGIng}) can be formulated as
\begin{eqnarray*}
g_c\left( x\right)  &=&\frac 1{2\pi }+\frac 1{2\pi \lambda
_s}\int_{-1}^1dy
\frac{g_s\left( y\right) }{1/4+\left( x/\lambda _c-y/\lambda _s\right) ^2} \\
g_s\left( y\right)  &=&\frac 1{2\pi \lambda _c}\int_{-1}^1dx\frac{g_c\left(
x\right) }{1/4+\left( y/\lambda _s-x/\lambda _c\right) ^2} \\
&&-\frac 1{\pi \lambda _s}\int_{-1}^1dy^{\prime }\frac{g_s\left( y^{\prime
}\right) }{1+\left( y-y^{\prime }\right) ^2/\lambda _s^2}
\end{eqnarray*}
with $\lambda _c=\gamma \int_{-1}^1dxg_c\left( x\right) $ and
$\lambda _s= \frac{2\gamma }{1-\zeta }\int_{-1}^1dyg_s\left(
y\right) $. After solving the integral equations the energy density
of ground state can be evaluated by
\begin{eqnarray*}
\varepsilon _{GS}=\frac{\hbar ^2N^2\gamma ^3}{2mL^2\lambda _c^3}
\int_{-1}^1x^2g_c\left( x\right) dx.
\end{eqnarray*}
For the system with total density $\rho =\rho _1+\rho _2$ and
polarization $ \zeta =\left( \rho _1-\rho _2\right) /\rho $, the
energy density of ground state can be parameterized as
\begin{equation}
\varepsilon _{GS}\left( \rho ,\zeta \right) =\frac{\pi ^2\hbar ^2\rho ^2}{8m}%
\left[ \left( 1/3+\zeta ^2\right) +f\left( \chi ,\zeta \right)
\right] \label{GSEFermi}
\end{equation}
with
\begin{eqnarray}
f\left( \chi ,\zeta \right)  &=&\left[ e\left( \chi \right) -1/3\right]
\left\{ 1+\alpha \left( \chi \right) \zeta ^2+\beta \left( \chi \right)
\zeta ^4\right. \nonumber \\
&&\left. -\left[ 1+\alpha \left( \chi \right) +\beta \left( \chi \right)
\right] \zeta ^6\right\} ,
\end{eqnarray}
where $\chi =2\gamma /\pi $, and $e\left( \chi \right) $, $\alpha \left(
\chi \right) $ and $\beta \left( \chi \right) $ are given in Ref. \cite
{Gao06, Gao07}
\begin{eqnarray}
e\left( \chi \right)  &=&\frac{4\chi ^2/3+a_p\chi +b_p}{\chi ^2+c_p\chi +d_p}%
,  \nonumber \\
\alpha \left( \chi \right)  &=&\frac{-\chi ^2+a_\alpha \chi +b_\alpha }{\chi
^2+c_\alpha \chi -b_\alpha },  \nonumber \\
\beta \left( \chi \right)  &=&\frac{a_\beta \chi }{\chi ^2+b_\beta \chi
+c_\beta }  \label{fermifitfunc}
\end{eqnarray}
with $a_p=5.780126$, $b_p=-(8/9)ln2+\pi a_p/4$, $c_p=(8/\pi )ln2+3a_p/4$, $%
d_p=3b_p$, $a_\alpha =-1.68894$, $b_\alpha =-8.0155$, $c_\alpha =2.74347$, $%
a_\beta =-1.51457$, $b_\beta =2.59864$ and $c_\beta =6.58046$.

Thus for the mixture of two-component TG gas, the energy density
$\epsilon (\rho )$ in energy functional (\ref{EF}) should be
replaced by $\epsilon _{GS}(\rho ,\zeta )|_{\rho \rightarrow \rho
\left( x\right) ,\zeta \rightarrow \zeta \left( x\right) }$
\cite{Zubarev,Gao06}. With the same procedure as the former section,
the coupled Schr\"{o}dinger equations take the same formula as Eq.
(\ref
{NLEs}) and $\tilde{F}\left( \rho \right) $ in the equation about $\Phi _i$ (%
$i=1,2$) should be replaced with
\begin{eqnarray}
\tilde{F}_i\left( \rho ,\zeta \right) =\frac{\pi ^2\hbar
^2}{8m}\frac
\partial {\partial \rho _i}\rho ^3\left[ \left( 1/3+\zeta ^2\right) +f\left(
\chi ,\zeta \right) \right].
\end{eqnarray}
For the unpolarized case, it is reduced to
\begin{eqnarray}
\tilde{F}\left( \rho \right) &=&\frac{\pi ^2\hbar ^2}{8m}\left[
3\rho ^2e\left( \chi \right) +\rho ^3 \frac{\partial e}{\partial
\rho }\right].
\end{eqnarray}
By numerically solving the coupled NLEs, we can obtain the
ground-state density distributions for different interacting
constants $U$ ($U =g/l\hbar \omega$).
\begin{figure}[tbp]
\includegraphics[width=3.2in]{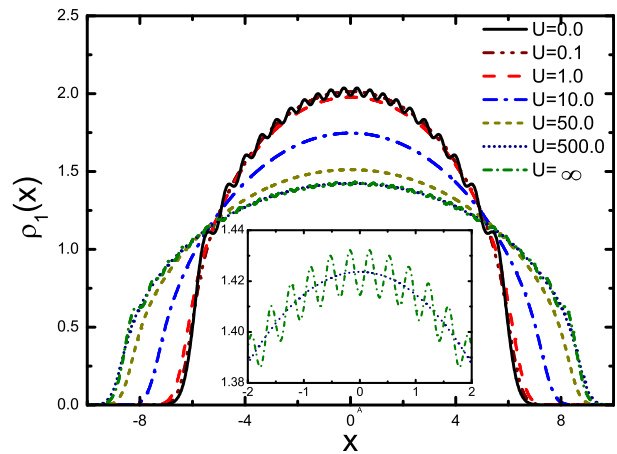}\newline
\caption{(color online) Ground state density distribution for
unpolarized two-component TG gases with $N_1=N_2=20$. Unit of
$\rho_1(x)$: $1/l$; Unit of $x$: $l$.} \label{fig4}
\end{figure}

We first consider the unpolarized Bose mixture with $N_1=N_2$, for
which we always have $\rho_1(x)=\rho_2(x)$. In Fig. 4, we show the
density profiles $\rho_1(x)$ for various inter-species interactions,
which are normalized to the atom number $N_1$. When the
inter-species interaction is zero, no correlation exists between two
TG gases and $\rho_1(x)$ is given by the density distribution of the
single-component TG gas, which is identical to the density
distribution of a polarized Fermi gas with $N_1$ fermions
\cite{Girardeau}. When the inter-species interaction is weak, the
density distribution of each component does not deviate far from the
distribution of the single-component TG gas. With the increase in
inter-species interaction, the density profiles become broader and
broader, and continuously evolve to the limit of strong
inter-species interaction. As shown in Fig.4, the density profile
for $U=500$ already agrees very well with the exact profile in the
limit of $U=\infty$ except the oscillation peaks. In this limit, the
density profile for each component fulfils the relation (\ref{ruo})
and displays $N$ peaks, which can be exactly calculated by using the
Bose-Fermi mapping method \cite{Girardeau07}.
\begin{figure}[tbp]
\includegraphics[width=3.5in]{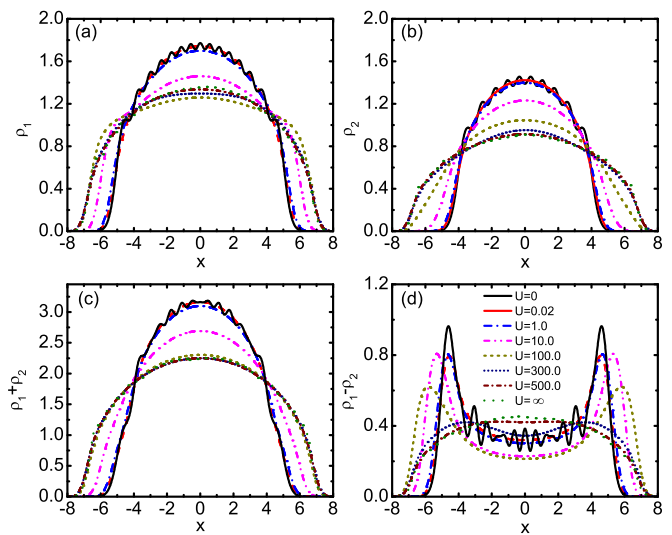}\newline
\caption{(color online) Ground state density distribution for
polarized two-component TG gases with $N_1=15$ and $N_2=10$. Unit of
$\rho_i(x)$: $1/l$; Unit of $x$: $l$.} \label{fig5}
\end{figure}

For the imbalance trapped mixture of TG gases, the distributions of
each component will display different behaviors in the full
interacting regime except in the limit of strong interspecies
interaction, where the distributions display same behaviors only
with different normalization condition, {\it i.e.}, $n_1(x)= n_2(x)$
with $n_i(x)= \rho_i(x)/N_i$. The density profiles of each
components and the total density profiles for the polarized mixture
of TG gases are exhibited in Fig. 5a, Fig. 5b and Fig. 5c,
respectively. The profiles for the case of vanished inter-species
interaction and of infinitely strong inter-species interaction are
also plotted (oscillating lines) in the figure, which are obtained
by means of BFM method. It is shown that in the regime close to
these two limits the distributions agree well with the results of
BFM method. It is feasible even in the situation of finite weak or
strong inter-species interaction. With the increase in inter-species
interaction both components distribute in more extensive regime and
the density profiles of $i$-component evolve from the behavior of
single-component TG gas with $\rho_i(x)=\rho_{TG}(x)|_{N=N_i}$ to
the behavior of the fermionized Bose mixture with  $\rho_i(x)=\frac
{N_i}{N} \rho_{TG}(x)|_{N=N_1+N_2}$. In Fig. 5d, we show the spin
density distribution defined as the density difference between two
components. For $U=0$, we have $\rho_1(x)-\rho_2(x)=
\sum_{n=N_2}^{N_1-1} |\phi_n(x)|^2  $. In the weakly and
intermediately interacting regimes, the spin density distributions
display two explicit peaks at the location far away from the center
of trap and at the locations near the middle of trap spin density
profiles are almost flat. That is to say, two weakly interacting TG
gases form the shell structure of the coexisted regime of two
components surrounded by the fully polarized majority component.
With the increase in interspecies interaction, the peaks become
smaller and smaller and disappear finally. In the limit of strong
inter-species interaction, two components coexist in the full spaces
and the spin density profile exhibits similar behavior to the total
density profile. This could be understood because in the
infinite-$U$ limit one has $\rho_1(x)-\rho_2(x)=\frac{N_1-N_2}{N}
\rho_{tot}(x)$ according to (\ref{ruo}).

\section{Summary}
In summary, the ground-state properties of two-component Bose gases
trapped in 1D harmonic traps have been studied in the scheme of DFT.
On the basis of Bethe-ansatz solution, we calculate the ground-state
density distributions of the confined systems for each component by
combining the local density approximation with the exact solution of
homogeneous system. For the case with spin-independent interaction,
the distribution of each component is $N_i /N$ of the total density
distribution in the full repulsive regime. With the increase in the
interaction the density distributions show evolution from Gauss-like
Bose distributions to Fermi-like distributions. For the mixture of
two TG gases with tunable inter-species interaction, we show that it
can be mapped into a two-component Fermi gas by a generalized
Bose-Fermi transformation. Taking advantage of the exact
ground-state energy of Yang-Gaudin model for the homogeneous system,
we calculate the density profiles of the confined systems.  For both
balanced Bose mixture and polarized Bose mixture, the density
profiles of each component display the continuum crossover from the
distribution of single-component TG gas with $N_i$ atoms to that of
composite-fermionization mixture as inter-species interaction
increases. In the limit of strong repulsion, the results obtained
from density-functional calculation are compared with the results
from the Bose-Fermi mapping method and it turns out that they match
very well in the full coordinate space.

\begin{acknowledgments}
This work was supported by NSF of China under Grants No. 10821403
and No. 10847105, programs of Chinese Academy of Sciences, and
National Program for Basic Research of MOST.
\end{acknowledgments}

\end{document}